\pdfoutput=1
\documentclass[english,prb,balancelastpage,groupedaddress,twocolumn,aps,10pt,floatfix]{revtex4-1}
\usepackage[T1]{fontenc}
\usepackage[latin9]{inputenc}
\usepackage{babel}
\usepackage[caption=false]{subfig}
\usepackage{amsmath}
\usepackage{amssymb}
\usepackage[resetlabels, labeled]{multibib}
\usepackage{graphicx}
\usepackage{dsfont}
\usepackage[unicode=true,pdfusetitle,
 bookmarks=true,bookmarksnumbered=false,bookmarksopen=false,
 breaklinks=false,pdfborder={0 0 1},backref=false,colorlinks=false]
 {hyperref}

\begin{document}

\title{Critical behavior of topological Kondo Insulators}

\author{J.J \surname{van den Broeke}}

\affiliation{Institute for Theoretical Physics, Center for Extreme Matter and
	Emergent Phenomena, Utrecht University, Princetonplein 5,
3584 CC Utrecht,
	The Netherlands}

\author{S.N. \surname{Kempkes}}

\affiliation{Institute for Theoretical Physics, Center for Extreme Matter and
	Emergent Phenomena, Utrecht University, Princetonplein 5,
3584 CC Utrecht,
	The Netherlands}

\author{A. \surname{Quelle}}

\affiliation{Institute for Theoretical Physics, Center for Extreme Matter and
	Emergent Phenomena, Utrecht University, Princetonplein 5,
3584 CC Utrecht,
	The Netherlands}

\author{C. \surname{Morais Smith}}

\affiliation{Institute for Theoretical Physics, Center for Extreme Matter and
	Emergent Phenomena, Utrecht University, Princetonplein 5,
3584 CC Utrecht,
	The Netherlands}

\date{\today}
\begin{abstract}
A thermodynamic study of the Kondo insulator SmB$_6$ is pursued to elucidate the well-known anomalous low-temperature electronic-like specific heat contribution conjectured to arise from metallic surface states. A general thermodynamic description of topological Kondo insulators is developed using a mean-field slave-boson approximation and an approach inspired by Hill thermodynamics to study the phase transitions with the critical exponents of the model. The results show consistency with the Josephson hyper-scaling relation. We further apply this framework to SmB$_6$. By separating the bulk and boundary contributions to the heat capacity, we show that while the surface states contribute to an increase of the heat capacity upon lowering the temperature, the effect is immeasurably small. This suggests that an alternate explanation is required for the anomalous low-temperature contributions to the specific heat in this material.

\end{abstract}
\maketitle

\section{Introduction}

Topological insulators are materials that are insulating in the bulk, but have conducting surface states at the boundaries. These surface states are protected by the symmetries of the bulk Hamiltonian against impurity scattering, and lead to fascinating phenomena, such as Majorana zero modes in 1D \cite{kitaev2001unpaired,mourik2012signatures} or the quantum spin Hall effect in 2D. \cite{kane2005quantum,yao2007spin,min2006intrinsic,bernevig2006quantum,konig2007quantum} In 3D, there are also several experimental realizations of topological insulators, such as Bi$_2$Se$_3$ and Bi$_2$Te$_3$,\cite{zhang2009topological,xia2009observation} but these materials have a residual bulk conductivity caused by impurities, \cite{moore2010birth} and are therefore not truly gapped.    

One of the most promising realizations of a 3D topological insulator with a fully gapped bulk is SmB$_6$.\cite{xu2016spin,pirie2020imaging,ohtsubo2019non,rosa2020bulk} This material is a Kondo insulator, i.e. below the Kondo temperature the otherwise metallic system becomes insulating. This transition is caused by a hybridization between conducting $d$-electrons and localized $f$-electrons that occurs at low temperatures, opening up a hybridization gap. In light of recent developments in the field of topological insulators, Dzero et al. \cite{dzero2010topological} proposed that Kondo insulators could have a topological phase. This would naturally explain the puzzling behavior of the residual conductivity of SmB$_6$ in the low-temperature (T) regime. In addition, another obscure feature of SmB$_6$ that baffled scientists for many years, namely the resemblance of the low-$T$ heat capacity to that of a metal, \cite{phelan2014correlation,nickerson1971physical} motivated the search for an explanation in the topological nature of the system. In this view, the upturn in the heat capacity would be ascribed to the presence of metallic edge states, but this hypothesis still lacks confirmation.

SmB$_6$ is usually described using the Anderson lattice model.\cite{anderson1961localized} In this tight-binding approach, an even number (spin up and spin down) of conducting $d$ and localized $f$ bands is considered. The different models available vary in the amount of bands that are included, the hybridization between the $f$ and $d$ bands, and in the range of hopping to which neighboring sites are incorporated. Although the minimal model for a Kondo insulator involves only four bands,\cite{dzero2010topological} it was proposed that a realistic description of SmB$_6$ requires a ten-band model, with up to third nearest-neighbor hopping.\cite{baruselli2014scanning} The parameters in this ten-band model are taken from ab-initio calculations,\cite{deng2013plutonium} and the resulting in-gap surface states around both the $\Gamma$ and $X$ high-symmetry points in the Brillouin zone agree well with angle-resolved photoemission spectroscopy measurements.\cite{xu2013surface,neupane2013surface,jiang2013observation,min2014importance,denlinger2014smb6}

However, the upturn in the heat capacity at low $T$ has not yet been explained using these tight-binding models, and it is unclear whether topological edge states can indeed solve the issue. One of the major difficulties is that topological systems have mostly been studied at zero $T$, whereas the heat capacity is a finite-$T$ property. Recently, a thermodynamic approach inspired by the pioneering work of Hill \cite{hill1963thermodynamics} has been proposed, which allows to separate the bulk and boundary contributions to the free energy, and describe the topological phase transition at zero and finite $T$. \cite{quelle2016thermodynamic} Moreover, it was shown that the order of the topological phase transition in the bulk and at the boundary obeys an universal law for the five most common models for topological insulators, namely the SSH and Kitaev in 1D, the Kane-Mele and BHZ in 2D and the BHZ in 3D.\cite{kempkes2016universalities,yunt2020topological} Although the phase diagram of Kondo insulators has been determined previously, \cite{dzero2012symplectic,werner2013interaction,pixley2015global} the order of the phase transition has not been considered. It is therefore interesting to verify whether the universal rule also applies to topological Kondo insulators.

In this work, we first determine the phase diagram of a topological Kondo insulator and investigate the phase transition from a topological to a band-insulator phase using the thermodynamic approach. For this, we employ a general minimal model to study the topological Kondo insulator, accounting for only two $d$ and two $f$ bands. We then compare our results to the previously found universality in other topological-insulator models,~\cite{kempkes2016universalities} and with the Josephson-hyperscaling relations. Second we examine SmB$_6$ in more detail by solving the $T$-dependent mean-field equations for a more realistic ten-band model. Then, we use an approach inspired by Hill thermodynamics to calculate the bulk and edge contribution to the heat capacity. We show that the edge contribution indeed yields an upturn, but this effect is not of the right order of magnitude to explain the experimental observations. This is in line with more recent experimental results.\cite{wakeham2016low} 

The outline of the paper is the following: in Sec.~\ref{sec:pt}, we briefly review the Anderson lattice model and the slave-boson mean-field approximation for the Kondo insulator. We use this model to analytically study the order of the topological phase transition, by taking derivatives of the free energy and looking for discontinuities. Then, we investigate the phase transition on a more general level, by determining the critical exponents for both bulk and boundary. In Sec.~\ref{sec:hetcap}, we calculate the heat capacity of SmB$_6$. Our conclusions and outlook are presented in Sec.~\ref{sec:ceno}.   

\section{\texorpdfstring{Topological phase transition in a Kondo insulator}{Topological phase transition in A Kondo insulator}} \label{sec:pt}
\subsection{The Model}
We begin by modeling the Kondo insulator using the well-known Anderson lattice model (ALM). \cite{hewson1997kondo,coleman1987mixed} The ALM  Hamiltonian consists of three parts,
\begin{equation}
H_{ALM}=H_d+H_f+H_h.
\end{equation}
Here, $H_d$ describes the conduction ($d$) electrons, $H_f$ the (almost) localized $f$-electrons, and $H_h$ the hybridization,
\begin{align} \label{eq:ham}
H_d=&\sum_{i\sigma l}^{}\epsilon^d_l d^\dagger_{i\sigma l}d_{i\sigma l}-\sum_{\left\langle ij\right\rangle\sigma ll' }^{}t^d_{ij\sigma ll'}(d^\dagger_{i\sigma l}d_{j\sigma l'}+h.c.), \nonumber\\
H_f=&\sum_{i s}^{}\epsilon^f_s f^\dagger_{i s}f_{i s}-\sum_{\left\langle ij\right\rangle s s' }^{}t^f_{ij s s'}(f^\dagger_{i s}f_{j s'}+h.c.)\nonumber\\ &+ U\sum_{i s s'}^{}f_{i s}^\dagger f_{i s}f_{i s'}^\dagger f_{i s'},\nonumber\\
H_h=&\sum_{i\sigma l}^{}\sum_{j s}^{}(V^{}_{i\sigma l,j s}d^{\dagger}_{i\sigma l}f_{j s}+h.c.),
\end{align}
where $d_{i\sigma l}^{\dagger}$ is the creation operator of a $d$-electron on site $i$ with spin $\sigma$ in orbital $l$, and $f_{j s}^\dagger$ creates an $f$-electron on site $j$ with pseudo spin $s$. Due to the presence of strong spin-orbit coupling, $s$ is not a spin index, but defined by the total angular momentum $J$, and its $z$-component $M$.\cite{dzero2012theory} Further, $\epsilon$ and $t$ are the on-site energy and nearest-neighbor hopping amplitude, respectively, $U>0$ is the strength of the interaction between the $f$-electrons, and $V^{}_{i\sigma l,j s}$ is the hybridization-matrix element.  

In order to obtain an effective model that is quadratic in the creation and annihilation operators, we use the slave-boson approximation, \cite{hewson1997kondo,coleman1987mixed,read1983solution} and take the limit $U\rightarrow \infty $. In the case of almost filled $f$-bands (the case we will consider later for SmB$_6$), this corresponds to projecting out the states with two or more $f$ holes per site. We apply this constraint using slave bosons. For this, we switch to the hole representation, and assume that the slave boson is filling up a site in the absence of an $f$-hole. Thus, we get the transformation
\begin{equation}\label{eq:trans}
f^\dagger_{s i}\rightarrow f^\dagger_{s i}b_i,
\end{equation}
where $b_i$ is the slave-boson annihilation operator. This enables us to formulate the constraint that projects out the doubly occupied states,
\begin{equation} \label{eq:constraint}
b_i^\dagger b_i+\sum_{s}^{}f_{is}^\dagger f_{is}=1.
\end{equation}
As a consequence, $b_i^\dagger b_i$ and $f_{is}^\dagger f_{is}$ cannot be simultaneously nonzero, and the slave bosons drop out for the on site energy,
\begin{equation}
f_{is}^\dagger b_i b_i^\dagger f_{is}=(1+b^\dagger_ib_i)f_{is}^\dagger f_{is}=f_{is}^\dagger f_{is}.
\end{equation}

We apply the transformation of the $f$-holes (Eq.~\ref{eq:trans}) to the Hamiltonian and impose the constraint (Eq.~\ref{eq:constraint}) using the Lagrange multipliers $\lambda_i$. Then, we perform a mean-field approximation, in which we replace the $b_i$, $b_i^\dagger$ operators by their expectation values $\langle b_i\rangle =\langle b_i^\dagger\rangle=b$, and $i\lambda_j$ by $\bar{\lambda}$. After implementing a Fourier transformation and dropping the constant terms, we obtain the effective Hamiltonian,
\begin{equation}\label{eq:heff}
H_{\rm eff}(k)=\left(
\begin{array}{cc}
-\epsilon_{ }^{d}\underbar{1}+t_{k}^{d}&- bV_{k}^{} \\
&\\
-bV_{k}^{\dagger} & -(\epsilon_{s}^f-\bar{\lambda})\underbar{1}+b^2t^f_{k}
\end{array}
\right), 
\end{equation}
 where $\underbar{1}$ is the identity matrix, accounting for the (pseudo) spin species. Thus, the mean-field effect of the interaction is a rescaling of the $f$-electron hopping and of the hybridization, and a shift in the on-site energy of the $f$-electrons. The corresponding mean-field equations for $b$ and $\bar{\lambda}$ are derived in appendix \ref{ap:dermf}, and read
\begin{align}\label{eq:bequation}
0&=b N_s\bar\lambda-\sum\limits_{k}^{} {\rm Tr} \left[ \left\langle f^\dagger_{k} d_{k}\right\rangle V_k\right] +
b\sum\limits_{k}^{}{\rm Tr}\left[ \left\langle f^\dagger_{k} f_{k}\right\rangle t^f_k\right],\\\label{eq:lequation}
1&=b^2+\frac{1}{N_s}\sum_{k}^{}{\rm{Tr}}\left\langle f^\dagger_{k} f_{k}\right\rangle,
\end{align}
where $N_s$ is the number of sites, and we have used the notation for the propagator
\begin{equation}
G(\tau,\tau,k,k) =\left(
\begin{array}{cc}
\left\langle d^\dagger_{k} d_{k}\right\rangle & \left\langle d^\dagger_{k} f_{k}\right\rangle \\
&\\

\left\langle f^\dagger_{k} d_{k}\right\rangle & \left\langle f^\dagger_{k} f_{k}\right\rangle
\end{array}
\right).
\end{equation}
We consider specifically the regime where the number of holes per site is equal to the number of conduction bands $N_d$. This gives the additional constraint 
\begin{equation}
N_d=\frac{1}{N_s}\sum_{k}^{}{\rm Tr}\left\langle f^\dagger_{k} f_{k}\right\rangle+\frac{1}{ N_s}\sum_{k}^{}{\rm{Tr}}\left\langle d^{\dagger}_{k} d_{k}^{}\right\rangle.
\end{equation} 
In practice, the use of this constraint may be avoided at low $T$, by guaranteeing that the chemical potential $\mu$ is inside the gap.

\begin{figure*}[t]
	\subfloat[][]{\includegraphics[width=0.3\linewidth]{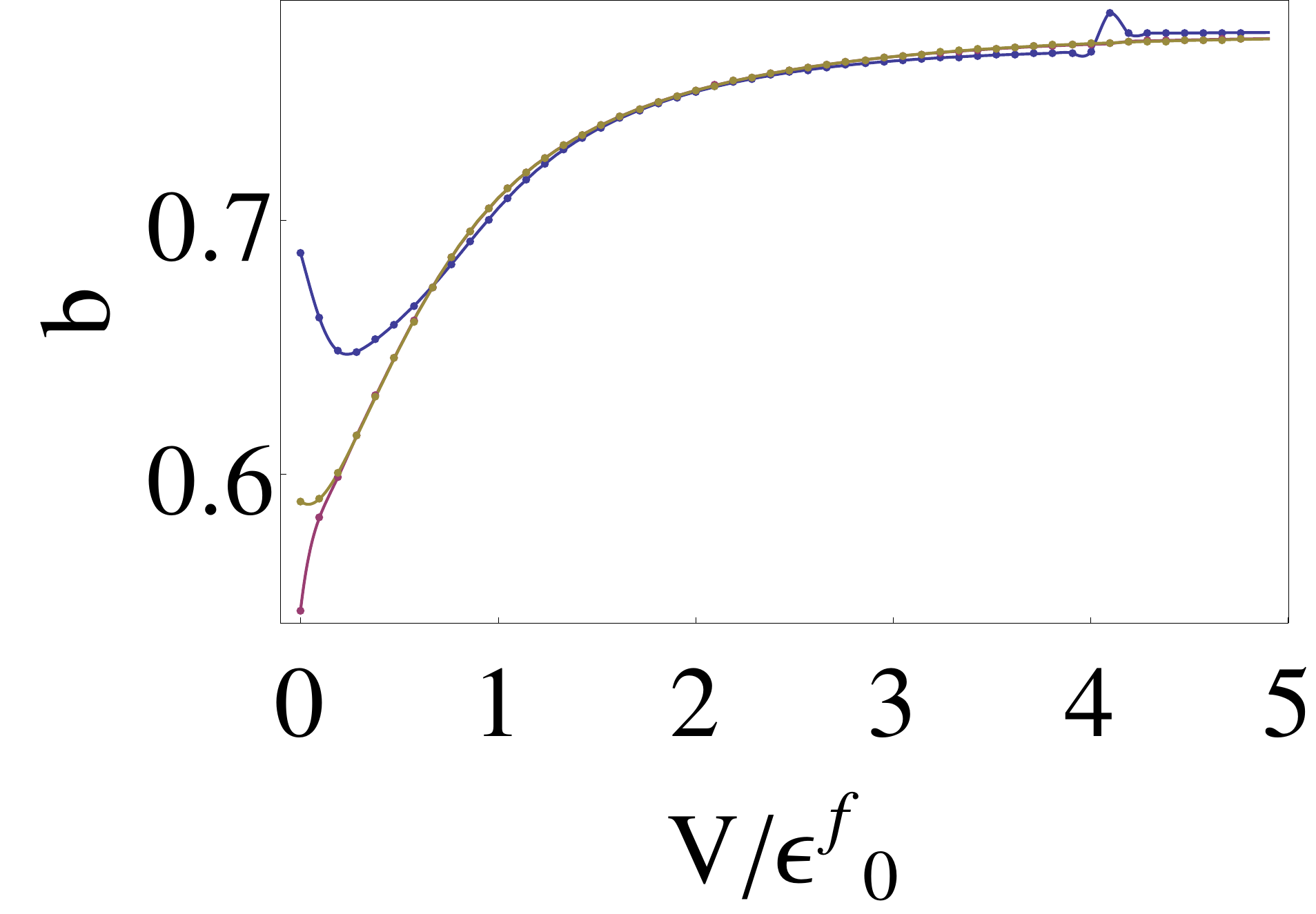}}
	\subfloat[][]{\includegraphics[width=0.28\linewidth]{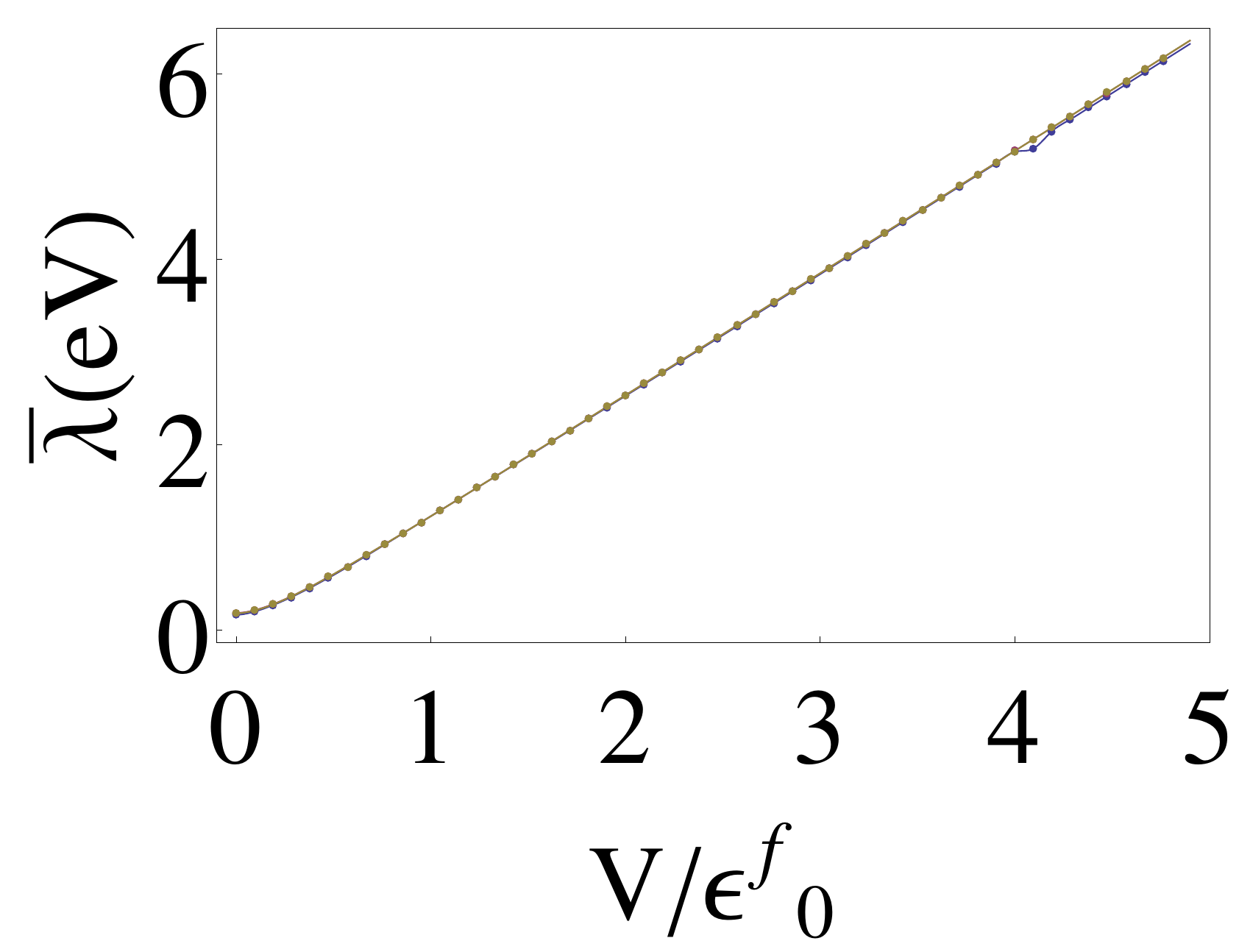}}
	\subfloat[][]{
		\includegraphics[width=0.29\linewidth]{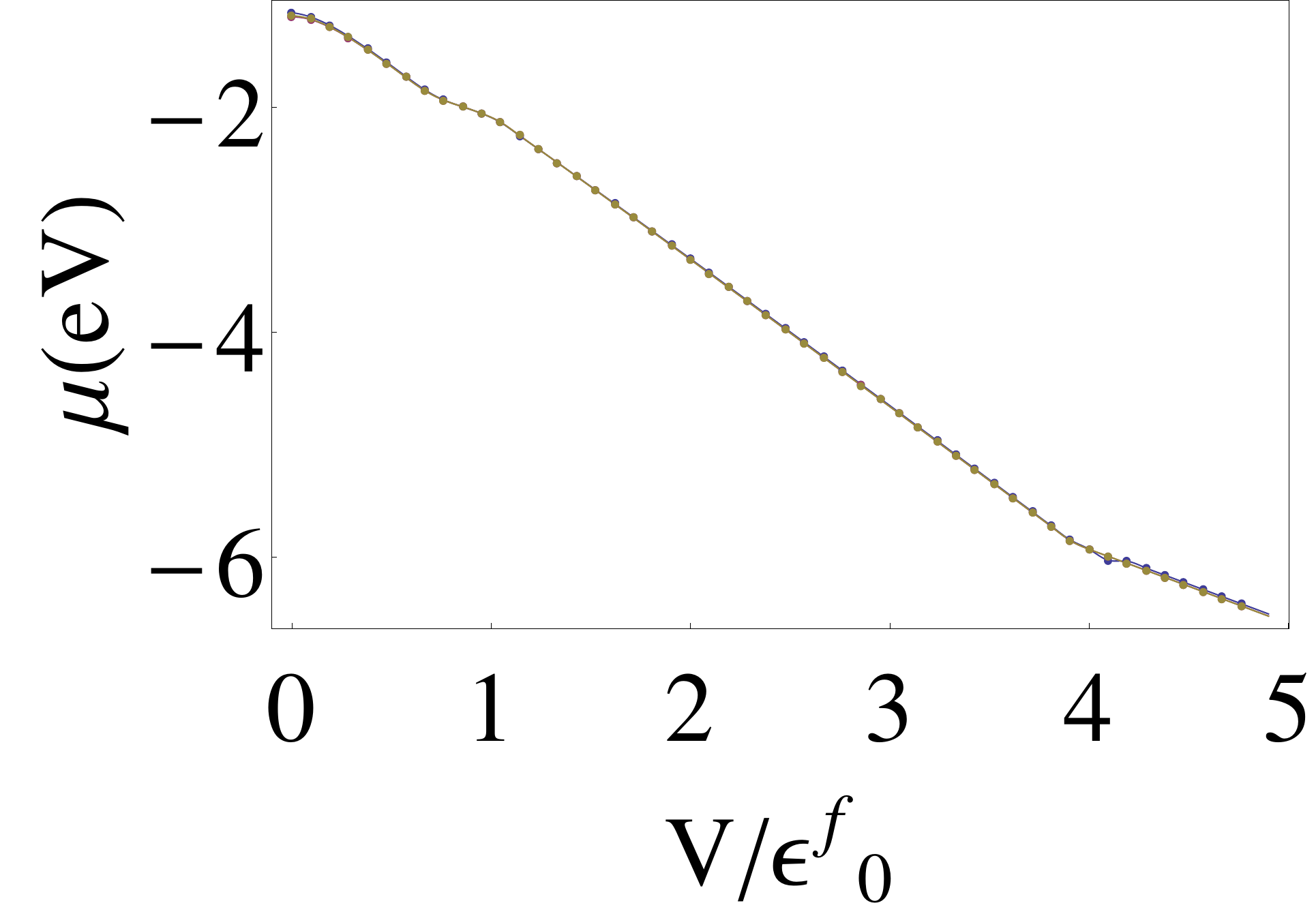}}
	
	\subfloat[][]{\includegraphics[width=0.3\linewidth]{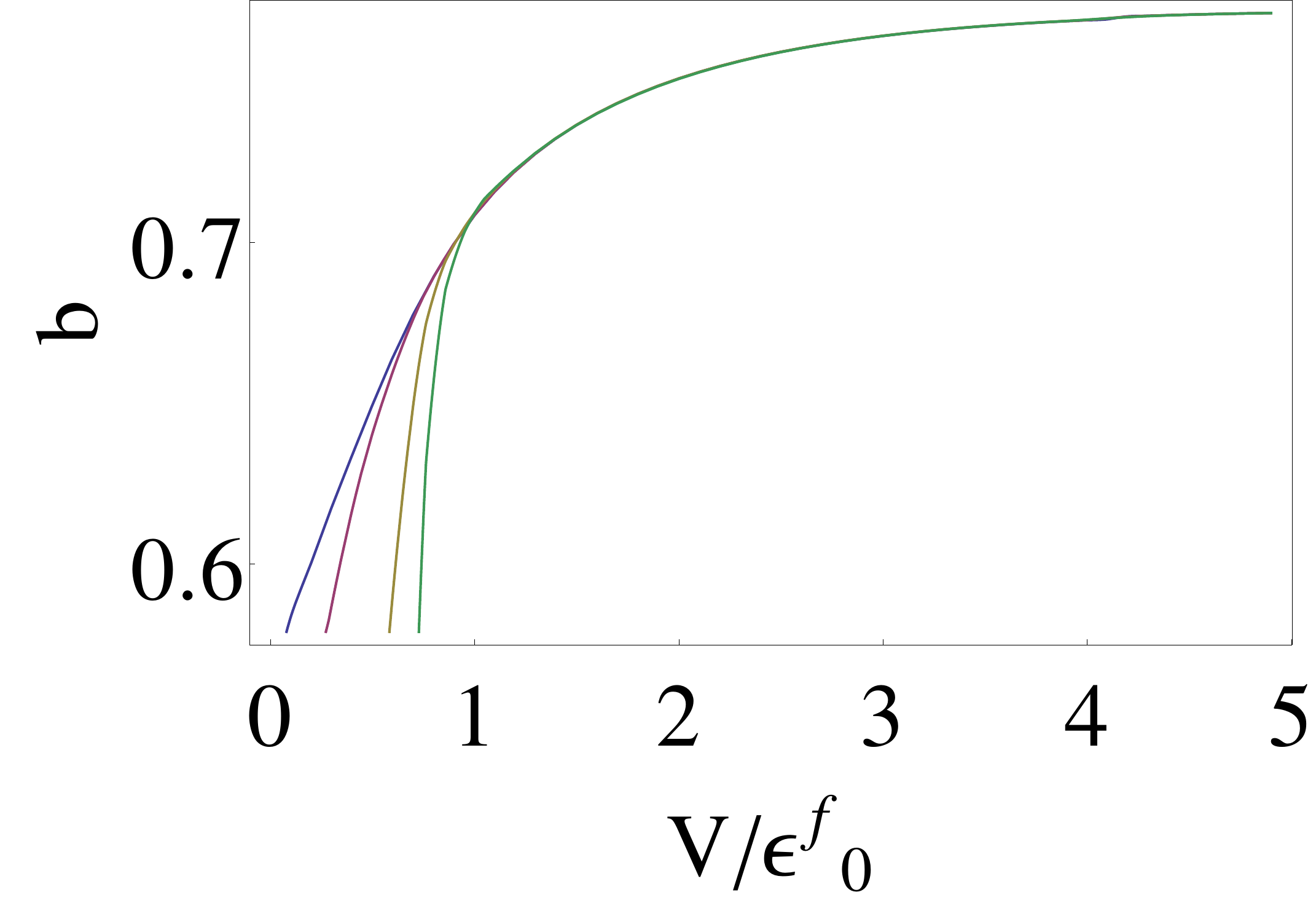}}
	\subfloat[][]{\includegraphics[width=0.28\linewidth]{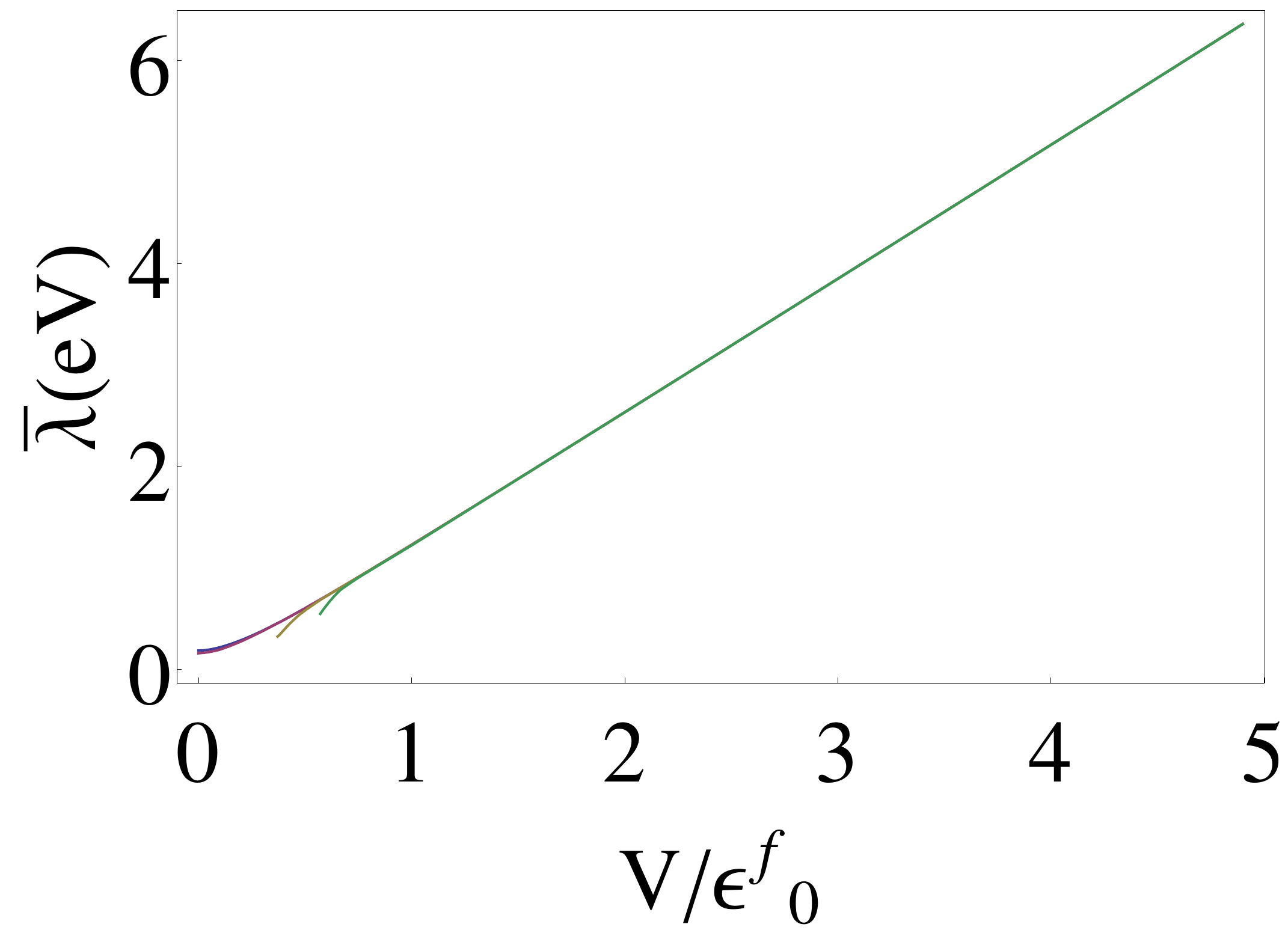}}
	\subfloat[][]{\includegraphics[width=0.29\linewidth]{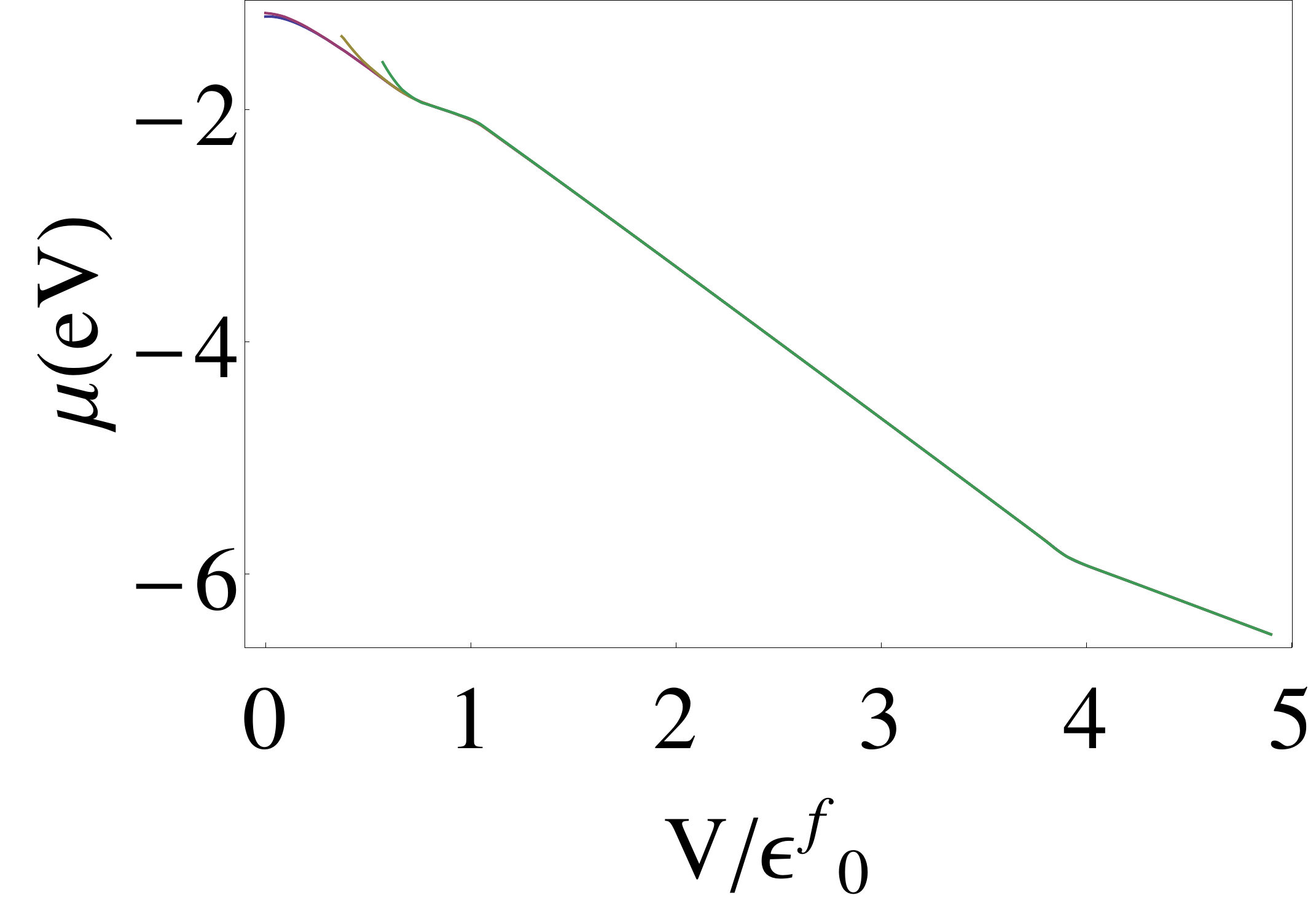}}

	\caption{Solutions to the mean-field equations as a function of hybridization. In (a),(b), and (c) blue, pink, and yellow indicate the solutions obtained using a $k$ space of respectively $5^3$, $11^3$, and $17^3$ points at zero $T$. In (d), (e), and (f)  blue, pink, yellow, and green indicate the solutions for $T/\epsilon_0^f$ equal to 0, 0.025, 0.049, and 0.074, respectively. Here, a $k$ space of $11^3$ points is used.}\label{fig:mfsol}
\end{figure*}

Now, we investigate the phase behavior of the minimal model that includes one $d$ and one $f$ orbital, and both spin up and down degrees of freedom. \cite{alexandrov2015kondo,dzero2010topological} In this case, 
\begin{align}\label{eq:4mod}
-\epsilon^d\underbar{1}+t^d_k&=-\underbar{1}\xi^d_k \nonumber \qquad
-\epsilon^f\underbar{1}+b^2t^f_k=-\underbar{1}\xi^f_k \nonumber\\
V_k&=V(\sin{k_x},\sin{k_y},\sin{k_z})^T.\overrightarrow{\sigma},
\end{align}
where \begin{align}\xi_k^d= -2t  \sum_{k_\eta}\cos{k_\eta}-\mu, \text{ and }\nonumber \\\xi_k^f= 2b^2t_f  \sum_{k_\eta}\cos{k_\eta}-\mu+\epsilon^f_0,
\end{align}
with $\eta=\{x,y,z\}$ and $t$ and $t_f$ are constant hopping parameters. Replacing Eq.~(\ref{eq:4mod}) into Eq.~(\ref{eq:heff}), and using the numerical values  $\epsilon^f_0=-1.05t$ and $t_f=0.1t$, we can solve the mean-field equations as a function of $T$ and hybridization $V$, which is outlined in appendix \ref{ap:dermf}. The results are shown in Fig.~\ref{fig:mfsol}. Here, we see that if the periodic $k$-space grid used to calculate the solutions is too small, the solution becomes unstable around $V/\epsilon_0^f=4.1$. As we will see later, this is exactly the point of the phase transition, where the correlation length diverges. Furthermore, kinks are visible in the $\mu$ solution. This is an artificial effect, caused by the approximation that $\mu$ is in the middle of the gap. Finally, we have that for $T/\epsilon_0^f>0.04$, $b$ becomes zero for finite $V/\epsilon_0^f$. This signals a phase transition from a Kondo insulator to a Kondo liquid phase, in which the mean-field equations no longer hold. 

Because the system is inversion symmetric, we use the parity eigenvalues at the high-symmetry points of the occupied bands to calculate the topological index $\nu$. \cite{fu2007topological} Specifically, we use the formula, 
\begin{equation}
(-1)^\nu=\prod_{i}^{}\prod_{m=1}^{N}p_{2m}(\Gamma_i),
\end{equation}
where $p_{2m}(\Gamma_i)=\pm1$ is the parity eigenvalue at the high-symmetry point $\Gamma_i$ of the $2m$'th energy band. Note that $p_{2m}=p_{2m-1}$, as this is a Kramers pair. The product involves all high-symmetry points in the Brillouin zone and the $2N$ occupied bands. If $\nu=1$, the system is a strong topological insulator, and the topological edge states are protected. If $\nu=0$, but one of the products involving $\Gamma_j'$s at high symmetry points in the same plane is $-1$ i.e, $\prod_{j}^{}\prod_{m=1}^{N}p_{2m}(\Gamma_j')=-1$, the system is a weak topological insulator (WTI). 

\begin{figure}[ht]
	\includegraphics[width=0.9\linewidth]{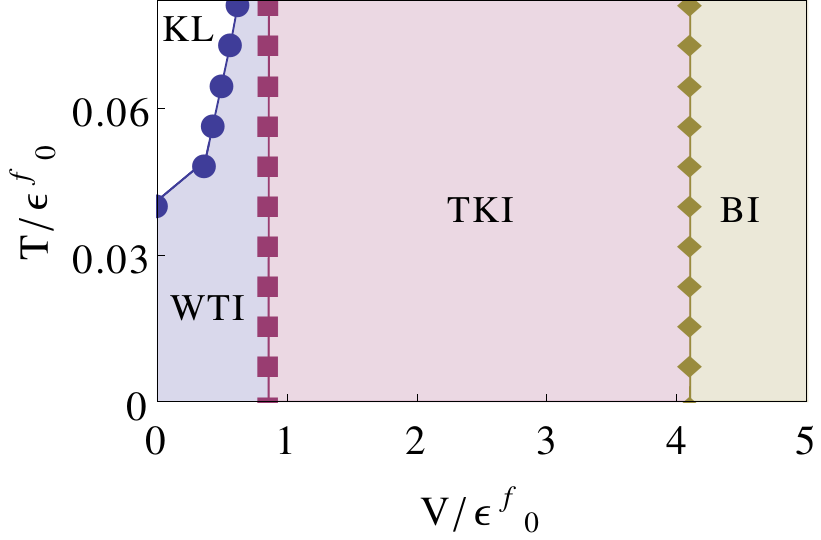}
	\caption{Temperature versus hybridization phase diagram of the four-band Kondo model. Four phases can be distinguished, a Kondo liquid (KL), a weak topological insulator (WTI), a topological Kondo insulator (TKI), and a band insulator (BI) phase.}
	\label{fig:pd}
\end{figure}
 The  resulting $T$ versus $V$ phase diagram for the four-band model is given in Fig.~\ref{fig:pd}. As promptly observed in the phase diagram, the topological phase transitions in this model are driven by the hybridization $V$ or by the temperature $T$. For low $V$ and $T$, the system is a WTI. As $V$ increases, the mean-field solution for $\bar\lambda$ increases, resulting in a relative shift of the $f$-electron energy, which changes the topological index (the $d$ and $f$ bands have an opposite parity eigenvalue), giving rise to a (strong) topological Kondo insulator (TKI) for intermediate $V$. If $V$ is increased even further, the $f$-bands move under the $d$-electron bands. As a consequence, there is no more band inversion and thus all parity eigenvalues at the high-symmetry points of the occupied bands at low $T$ become the same, resulting in a trivial topological index. The system then becomes a band insulator (BI). For sufficiently high $T$ and low $V$, the gap closes and the system becomes a Kondo liquid (KL). It should be noted that the phase diagram is strongly dependent on the number of bands. As shown by Dzero et al,\cite{dzero2012symplectic} inclusion of more bands leads to a disappearance of the WTI phase and an expansion of the stability region of the KL phase. 
\subsection{Free Energy}
\begin{figure}[b]
	\includegraphics[width=0.8\linewidth]{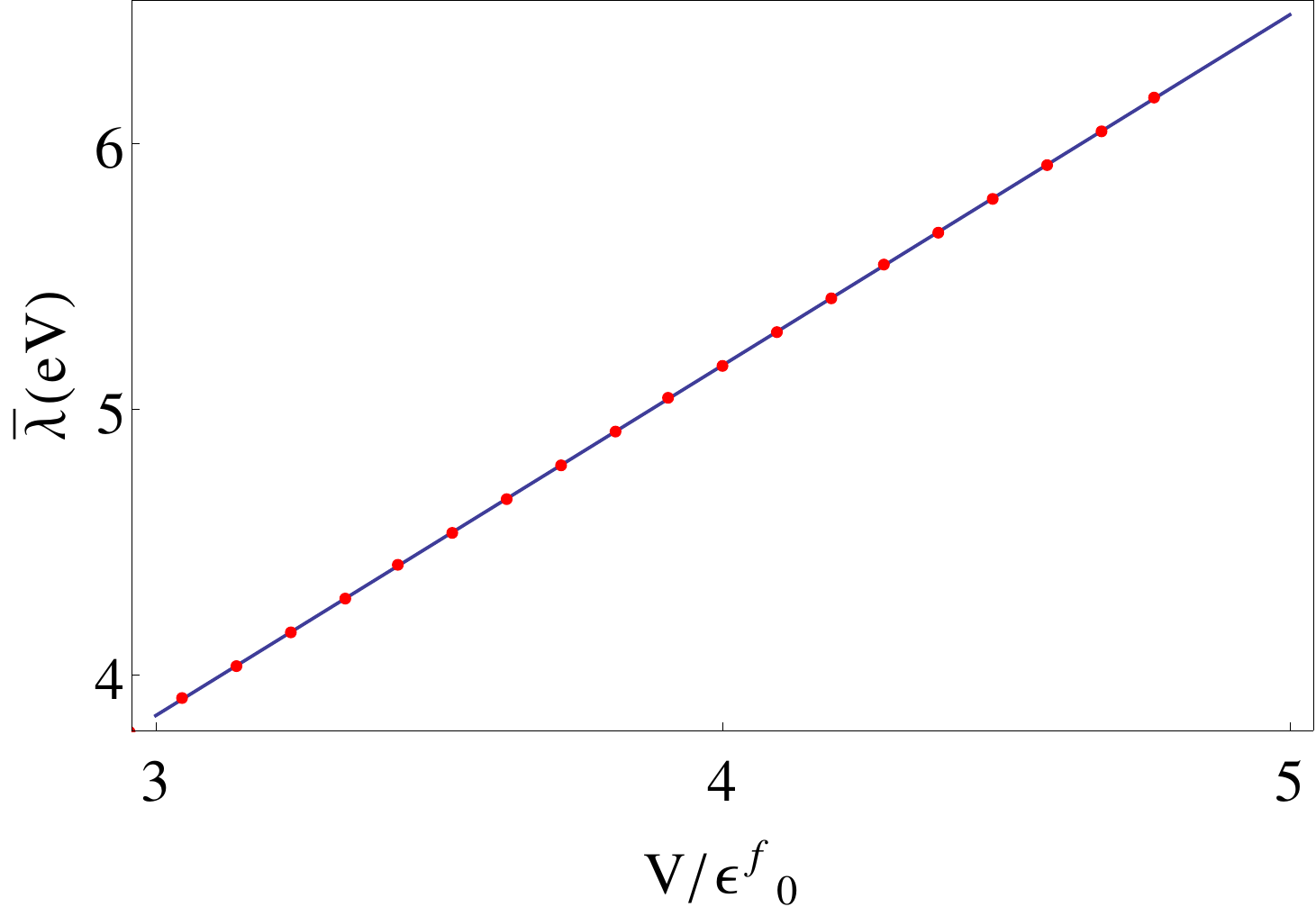}
	\caption{Numerical mean-field solution for $\bar\lambda$ at $T=0$ in the four-band model. The dots indicate the numerical solutions, whereas the blue line represents a linear fit.}
	\label{fig:lambda}
\end{figure}

To gain more insight in these phases and their transitions, we focus on the phase transition from a TKI to a BI and analytically determine its order for the bulk transition. The eigenvalues of the four-band Hamiltonian follow from diagonalizing Eq.~(\ref{eq:heff}),
\begin{multline}
E_{k\pm}=\frac{1}{2}(\xi_k^d+\xi_k^f-\bar{\lambda})  \\\pm \sqrt{(\xi_k^d-\xi_k^f+\bar{\lambda})^2 +4bV\sum_{k,\eta=x,y,z}\sin{k_\eta}}.
\end{multline}
 As the gap closes around the $\Gamma$ point, the free-energy contributions in the low-$k$ limit are the relevant contributions for the thermodynamic behavior. In this limit, 
\begin{multline}
E_{k\pm}=(t-b^2t_f)k +\delta_\mu  \\\pm \sqrt{\delta_{\bar{\lambda}} +(2(b^2t_f+t)\delta_{\bar{\lambda}} +4bV)k}.
\end{multline}
Here, $\delta_{\mu}=\mu-\mu_c $ and  $\delta_{\bar{\lambda}}=\bar{\lambda}-\bar{\lambda}_c$, where the subindex $c$ stands for the value of the parameter at the critical point. This results in the zero-$T$ free energy 
\begin{align}
\mathcal{F}=&\int_{k<\Lambda} E_{k-}k dk\nonumber\\
=& G_1(V)+\frac{1}{8}\delta_{\bar{\lambda}}^4(V)G_2(V)^{-3}\sinh^{-1}\left( \frac{G_2(V)}{\left| \delta_{\bar{\lambda}}(V)\right|} \right) ,
\end{align}
where $G_1$ and $G_2$ are smooth functions near the phase transition and $\Lambda$ is the cutoff enforcing the small-$k$ approximation. By solving the mean-field Eqs.~(\ref{eq:bequation}) and (\ref{eq:lequation}) numerically, we find that $\bar{\lambda}$ scales linearly with $V$ near the phase transition (see Fig.~\ref{fig:lambda}, where the red dots denote the numerical solution and the blue line is a linear fit to the numerical data). According to the Ehrenfest classification, the order of the phase transition is determined by the derivative of the free energy  that exhibits a discontinuity or a divergence. Thus, since $\delta_{\bar{\lambda}}$ is zero at the critical point, we expect a fourth-order bulk phase transition for this system. The fourth-order value for the transition indeed agrees with the universality for other topological models  unveiled previously, where the bulk topological phase transition was of order $d+1$, with $d$ the dimensionality of the model.  \cite{kempkes2016universalities}
\subsection{Critical Exponents}
\begin{figure}[tbp]
	\subfloat[][]{\includegraphics[width=0.5\linewidth]{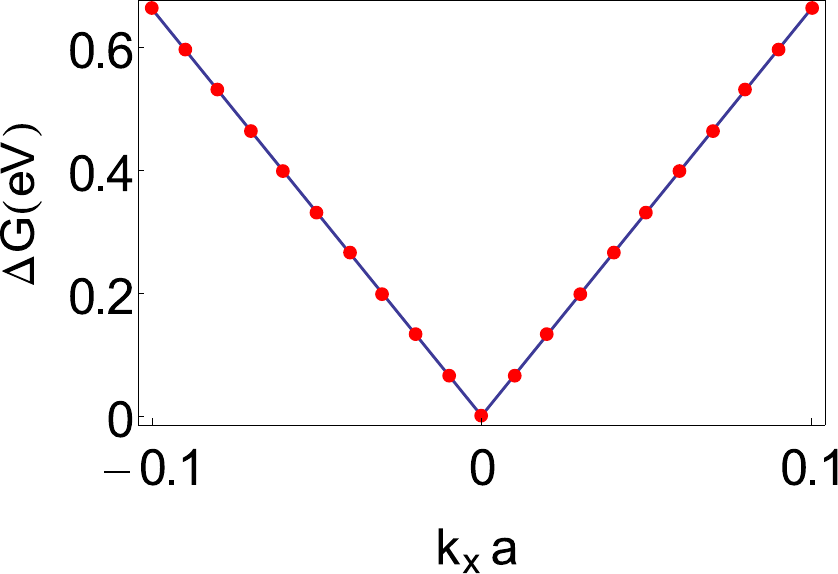}\label{fig:fluxa}}
       \subfloat[][]{\includegraphics[width=0.5\linewidth]{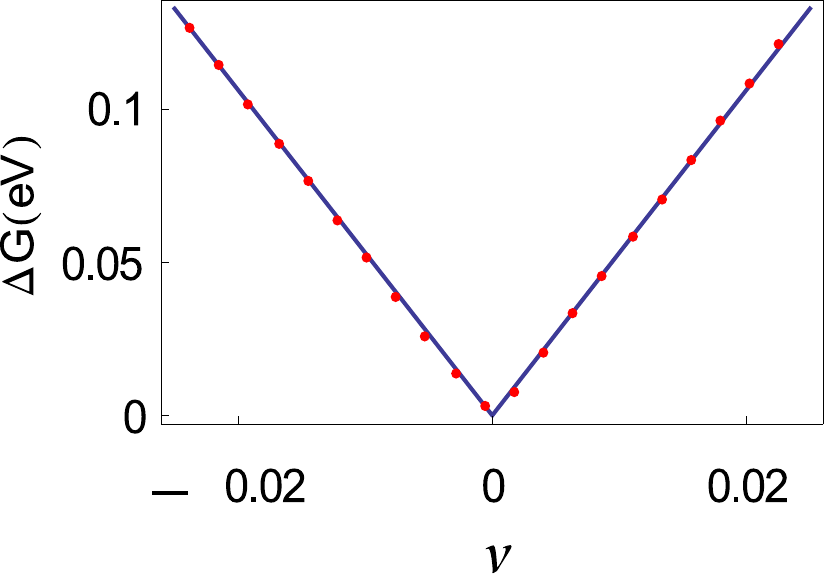}\label{fig:fluxa2}}
     \caption{Closing of the gap. The dots indicate the calculated gap size $\Delta G$ as a function of (a) the momentum $k_x$ for $k_y=k_z=0$ and (b) the reduced hybridization $v$. The blue lines indicate a linear fit to the numerical data and $a$ is the lattice constant. }
\label{fig:critexp}
\end{figure}

We can understand this universality in a more general way. Near the critical point, the singular part of the free energy of the system $\mathcal{F}$ scales with the dimensionless measure of the distance $v$ from the critical point,\cite{altland2010condensed}
\begin{equation}
\mathcal{F}(v)\propto\left| v\right|^{2-\alpha} . 
\end{equation}
In the case of a hybridization-driven transition, $v$ is equal to the reduced hybridization $v =\left| V-V_c\right|  /V_c$, where $V_c$ is the hybridization at the critical point. The quantum Josephson hyperscaling relation connects the order of the phase transition, which is equal to $2-\alpha$ in the Ehrenfest classification, to the critical exponents $\nu$ and $z$, \cite{josephson1966relation,continentino1989critical}
\begin{equation}
2-\alpha=\nu(d+z).
\end{equation}
Here, $d$ is the dimensionality of the system, and $\nu$ and $z$ can be determined by investigating the behavior of the energy gap $\Delta G$ near the critical point. \cite{continentino2017quantum} By considering $k_y=k_z=0$ and redefining $k_x=k$, we find
\begin{align}
\Delta G(V=V_c)\propto k^z,&  &
\Delta G(k=0)\propto \left|v \right| ^{\nu z}.
\end{align}
Next, we calculate how the gap closes as a function of $k$ and $v$, and find a linear behavior (see Fig.~\ref{fig:critexp}). Thus, we have $\nu=z=1$, which leads to the order of the bulk phase transition $2-\alpha=d+1=4$ in our case. This result puts topological Kondo insulators in the same universality class as the models investigated by Kempkes et al., \cite{kempkes2016universalities} which also had $\nu=z=1$.

\subsection{Behavior at the Edge}
\begin{figure}[ht]
	\includegraphics[width=0.8\linewidth]{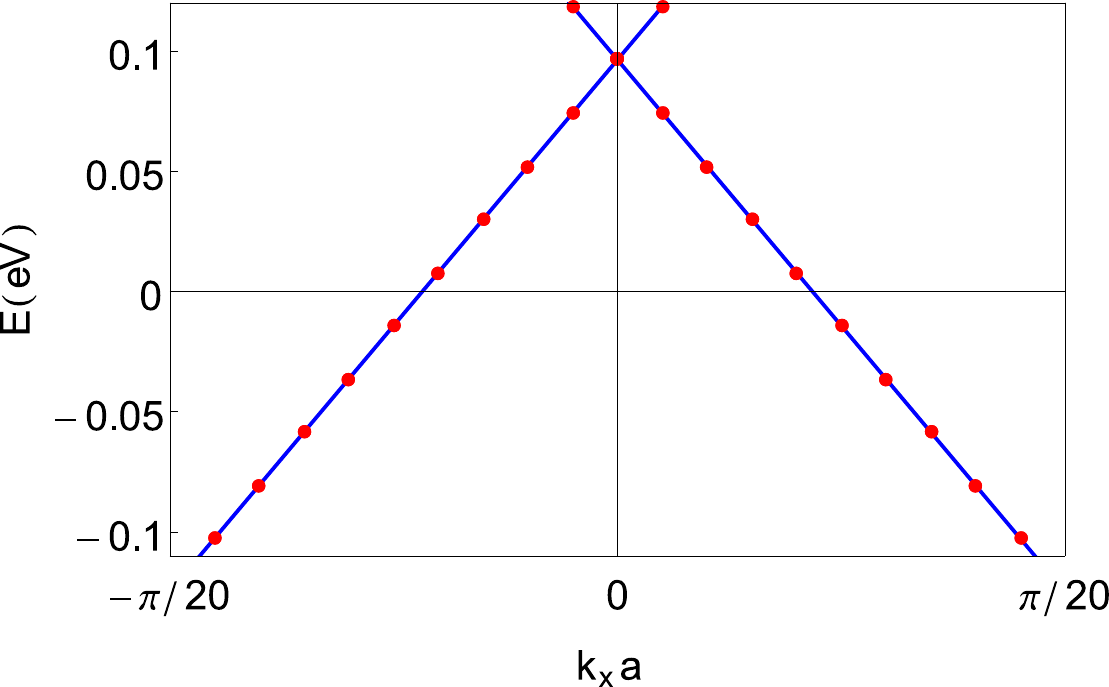}
	\caption{Edge state dispersion in the four-band model. The dots indicate the numerical calculations, the lines represent a linear fit to the data, and $a$ is the lattice constant. The dispersion was calculated for $V/\epsilon_0^f=3.8$, using a 400-layer system.  }
	\label{fig:edgev4}
\end{figure}

At the boundary, $\mathcal{F}$ scales with the critical exponents $\nu$ and $z'$, where $z'$ characterizes the dispersion of the edge states,
\begin{equation}
E(k)\propto k^{z'}.
\end{equation}  
Hence, the thermodynamic behavior of the boundary is dominated by the edge states that live in the gap, and we have
\begin{align}
\mathcal{F}&\propto\int_{\left| E(k)\right| <\Delta}^{}E(k)dk^{d-1}
\propto\int_{\left| k^{z'} \right| <\Delta}^{}k^{z'+d-2}dk \nonumber\\
&\propto\Delta^{\frac{1}{z'}(z'+d-1)}
\propto \left|v \right| ^{\frac{\nu z}{z'}(z'+d-1)},
\end{align}
where $\Delta=\Delta G(k=0)$ is the gap size. For the four-band topological Kondo insulator, we also find that $z'=1$, as shown in Fig.~\ref{fig:edgev4}. This indeed reproduces the universality for the boundary $2-\alpha=\nu(z'+d-1)$ found by Kempkes et al. \cite{kempkes2016universalities} for systems with $\nu=z'=1$. Hence, $\mathcal{F}\propto\left|v \right|^3$ and the phase transition at the boundary is of third order (a discontinuity occurs in the third derivative of the free energy).

\section{Heat capacity of SmB$_6$} \label{sec:hetcap}
\begin{figure}[ht]
	\includegraphics[width=0.9\linewidth]{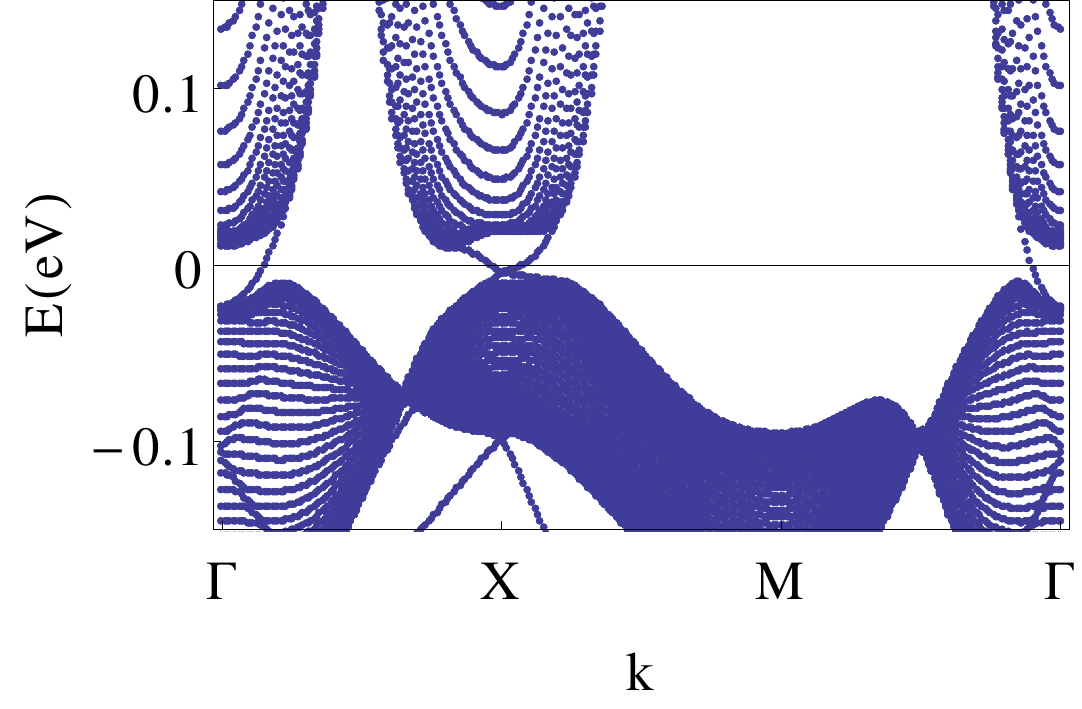}
	\caption{Finite-system band structure with 40 layers of the minimal model from ref.~\onlinecite{baruselli2014scanning}, where all first- and second-nearest neighbor hybridization terms are included.}
	\label{fig:finband}
\end{figure}

We now focus on a specific example of a Kondo insulator, namely SmB$_6$. The low-$T$ heat capacity of this material has long been known to present anomalous properties for a bulk insulator, namely it exhibits an upturn reminiscent of a metal. In order to investigate whether topological edge states can be responsible for this puzzling feature, we use the minimal model introduced in Ref.~\onlinecite{baruselli2014scanning}, which applies specifically for SmB$_6$. Here, we also include all first and second-nearest neighbor hybridization terms. This ten-band model, with the finite-system band structure shown in Fig. \ref{fig:finband}, includes the most important contributions of a full model with up to seventh nearest-neighbor hopping and hybridization, and reproduces the features observed experimentally,\cite{xu2013surface,neupane2013surface,jiang2013observation,min2014importance,denlinger2014smb6} as e.g. the in-gap edge states around the $\Gamma$ and $X$ points in the Brillouin zone.   

\begin{figure}[ht]
 	\includegraphics[width=0.9\linewidth]{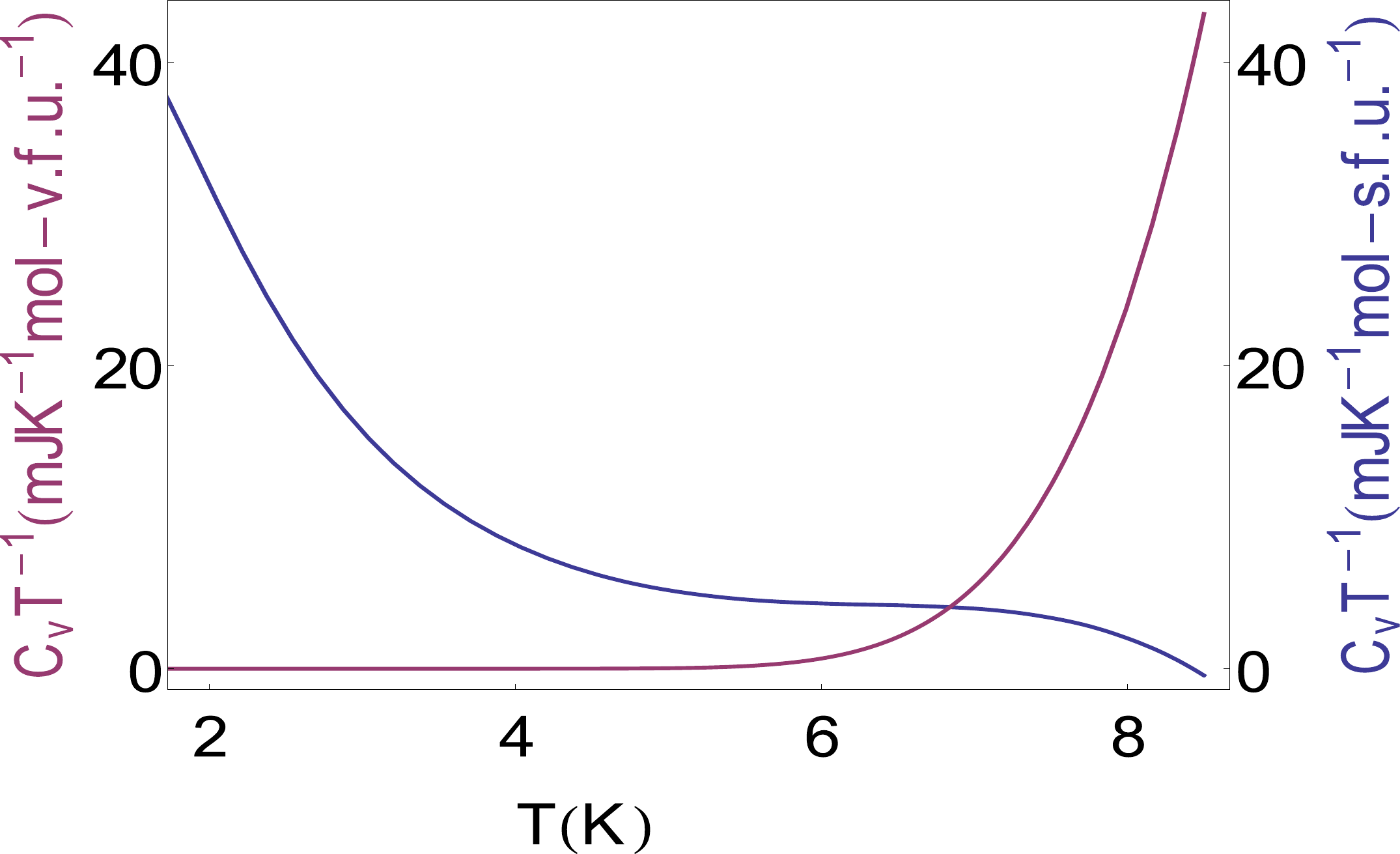}
 	\caption{Bulk (pink) and surface (blue) contribution to the heat capacity of SmB$_6$ described by the minimal model from Baruselli et al., \cite{baruselli2014scanning} where all first- and second-nearest neighbor hybridizations are included. Here, v.(s.)f.u. is an abbreviation for volume (surface) formula unit. The Kondo temperature is well above this regime.  }
 	\label{fig:hetcap}
 \end{figure}

Generally, thermodynamic quantities like the heat capacity are connected to the bulk of a system. In the case of topological insulators, however, the boundaries cannot be neglected due to the presence of edge states. This means that the conventional thermodynamic framework in which all quantities scale extensively cannot be used to describe topological insulators. We solve this problem by including the finite-size of a system in the thermodynamic quantities that allows for contributions to the grand potential $\Phi$ that do not scale extensively with the volume, but also with the surface or length, which is an approach inspired by Hill thermodynamics. \cite{hill1963thermodynamics} In this case, we expect the boundary modes to scale with the area, which leads to the following Ansatz for the grand potential
\begin{equation}\label{eq:hill}
\Phi=\mathcal{V}\Phi_\mathcal{V}+A\Phi_A,
\end{equation}
where $\mathcal{V}$ is the volume, $A$ is the surface area, and $\Phi_{\mathcal{V}(A)}$ is the bulk (surface) grand potential. It was recently shown that this formalism can be applied to describe the thermodynamic behavior of topological edge states.\cite{quelle2016thermodynamic, kempkes2016universalities, Cats}  Here, we firstly numerically solve the $T$-dependent mean-field equations for the ten-band model as described in appendix \ref{ap:dermf}, where the Hamiltonian is now a ten instead of a four-dimensional matrix. Then, we apply these solutions to a model with periodic boundary conditions in all but one direction. In the non-periodic direction the system consists of $n$ layers. If $n$ is large enough, such that there is no bulk-mediated interaction between the upper an lower layer,  Eq.~(\ref{eq:hill}) is valid. In this case, the grand potential can be separated into a boundary and a bulk contribution
 \begin{equation}\label{eq:hillsp}
 \Phi(T,n,l)=\Phi_{\mathcal{V}}(T)nl^2+\Phi_A(T)l^2,
 \end{equation} 
 where $l$ is the number of sites in the periodic directions. Using
 \begin{align}
 \Phi(T,n,l)=\frac{1}{\beta}\log\left[{\rm Tr}\left({\rm{e}}^{-\beta H_\text{eff}}\right)\right]
 =\frac{1}{\beta} \sum\limits_{j,k}^{} \log \left( 1+{\rm{e}}^{-\beta \epsilon_{jk}}\right) ,
 \end{align}
 where $\epsilon_{jk}$ is the $j$th eigenvalue of $H_{\rm eff}(k)$, given in Eq.~\ref{eq:heff}, we calculate the total grand potential, and then determine $\Phi_A(T)$ and $\Phi_\mathcal{V}(T)$ by making a linear fit to Eq.~(\ref{eq:hillsp}) for several values of $n$. Recalling that the heat capacity is equal to $-T\partial^2 \Phi_i/\partial T^2$, this procedure leads to the results shown in Fig.~\ref{fig:hetcap}, where we separated the bulk (pink) and surface (blue) contribution.  
 
 When we analyze Fig.~\ref{fig:hetcap}, we observe two features. First, the bulk contribution drops to zero for low $T$, which should indeed be the case for an insulator. Second, due to the metallic edge states, there is a peak in the heat capacity at low $T$, precisely as the one observed in the experiments, which go down to $2$K. However, this quantity is calculated per surface area. Upon computing the total heat capacity for the  experimentally used sample sizes, which are of the order of mm and thus contain approximately $ 10^7 $ atomic layers, we see that the upturn caused by the edges is seven orders of magnitude too small to explain the  experimentally measured heat capacity. \cite{phelan2014correlation}

\section{Conclusion and Outlook}\label{sec:ceno}
We investigated the thermodynamics of a Kondo system using both a simple four-band model to study the topological phase transitions, and a more detailed ten-band model to compute the low-$T$ heat capacity of SmB$_6$. In both cases, we used a slave-boson mean-field approximation. We found that Kondo insulators obey the same universality rule as the five most common topological insulator models, namely, the phase transition at the edge is one order lower than at the bulk.\cite{kempkes2016universalities} Furthermore, we see that this universal behavior can be understood by considering the critical exponents of the system. When calculating the critical exponents, care should be taken to select the right parameter range: as $V_c$ is calculated with numerical precision, $\Delta G(V=V_c,k=0)$ will only be zero up to the same precision. 

 Based on our calculations for SmB$_6$, we conclude that topological edge states are {\it{not}} responsible for the puzzling upturn in the heat capacity. This is in line with recent experimental results, where the heat capacity of powdered SmB$_6$ was compared to that of a single crystal sample and was shown to be similar.\cite{wakeham2016low} The idea that the anomalous heat capacity in SmB$_6$ is not a boundary effect is further supported by recent work showing a relation between the heat capacity upturn and bulk impurities in the material. \cite{thomas2019quantum,fuhrman2018screened,valentine2018effect,PhysRevLett.121.026602} Alternatively, it has been shown that the low temperature heat capacity behaviour can be explained by considering bulk excitons. \cite{knolle2017excitons} According to our results, although the edge states provide a qualitative upturn in the heat capacity, it is nearly impossible to measure this effect, as it is extremely small compared to experimentally observed upturn, even for a very thin sample of only a 100 atomic layers.


\section*{Acknowledgments} We would like to thank J. Paglione for his very helpful suggestions. Furthermore, we are grateful to P. Coleman, M. Dzero, J. Knolle, and M. Vojta for fruitful discussions. Financial support from the Netherlands Organization for Scientific Research (NWO) is gratefully acknowledged. The work by A.Q. and C.M.S. is part of the D-ITP consortium, a program of NWO that is funded by the Dutch Ministry of Education, Culture, and Science.   

\let\oldaddcontentsline\addcontentsline
\renewcommand{\addcontentsline}[3]{}
\bibliographystyle{apsrev4-1} 
\bibliography{p-wave,refs}  
\let\addcontentsline\oldaddcontentsline

\newpage
\widetext
\appendix
\setcounter{equation}{0}
\setcounter{figure}{0}
\setcounter{table}{0}
\setcounter{page}{1}
\makeatletter
\renewcommand{\theequation}{A\arabic{equation}}
\renewcommand{\thefigure}{A\arabic{figure}}
\renewcommand{\bibnumfmt}[1]{[A#1]}
\renewcommand{\citenumfont}[1]{A#1}
\renewcommand{\citenumfont}[1]{A#1}

\section{Derivation of the mean-field equations} \label{ap:dermf}
Here, we will show in detail how to derive and solve the mean-field equations (\ref{eq:bequation}) and (\ref{eq:lequation}) from the main text. Let us rewrite the effective hole Hamiltonian (\ref{eq:heff}), but keep track of the constant term resulting from the mean-field approximation, 
\begin{align}
H_{\rm eff}=&-\sum_{k\sigma l}^{}\epsilon_{ l}^{d} d_{k\sigma l}^{\dagger} d_{k\sigma l}^{}+\sum_{k\sigma ll'}^{}t_{k ll'}^{d} d_{k\sigma l}^{\dagger} d_{k\sigma l'}^{}\nonumber\\&-
\sum_{ks}^{}(\epsilon_{s}^f-\bar{\lambda})f^\dagger_{ks}f_{ks}
+b^2\sum_{kss'}^{}t^f_{kss'}f^\dagger_{ks}f_{ks'} \nonumber\\&-b\sum\limits_{ks\sigma l}^{}(V_{k\sigma s l}^{}f^\dagger_{ks}d_{k\sigma l}^{}+ H.c.) +  N_s(b^2-1)\bar\lambda,
\end{align}
where $N_s$ is the number of sites. In order to proceed, we need to calculate the values of $b$ and $\bar{\lambda}$. This can be done by minimizing the free energy associated with the effective Hamiltonian. The Euclidean action $S_{\rm eff}$ corresponding to $H_{\rm eff}$ is given by
\begin{equation}\label{seff}
S_{\rm eff}=
\hbar\beta N_s(b^2-1)\bar{\lambda}
-\int_{0}^{\hbar\beta}d\tau\sum_{k}^{}\int_{0}^{\hbar\beta}d\tau'\sum_{k'}^{} \Psi^\dagger(\tau,k) G^{-1}(\tau,k,\tau',k')\Psi(\tau',k'),
\end{equation}
where $\Psi(\tau, k)$ denotes the fermionic coherent state, and the inverse Green's function $G^{-1}$ reads 
\begin{equation}
G^{-1}(\tau,k,\tau',k')=\left(
\begin{array}{cc}
-\partial_\tau \underbar{1}+\epsilon^d_k -t^d_k& b{V}_k^\dagger 
\\ b{V}_k & -\partial_\tau\underbar{1}+ (\epsilon^f_k-\bar\lambda\underbar{1}) -b^2t^f_{k}
\end{array}
\right)\delta(k-k')\delta(\tau-\tau'),
\end{equation}
with $\underbar{1}$ denoting the identity matrix. This yields the partition function
\begin{equation}\label{Zequation}
Z={\rm Tr}\left( {\rm{e}}^{-\beta H_{\rm eff}}\right)	 =\int\limits_{\Psi(0)=-\Psi(\hbar\beta)}^{}\mathcal{D}(\bar{\Psi},\Psi)e^{-S_{\rm eff}/\hbar}=\exp\left[ -\beta N_s(b^2-1)\bar{\lambda}+{\rm Tr}\log(-G^{-1})/\hbar\right].
\end{equation}  
Next, the free energy $F=-k_BT\log{Z}$ is given by
\begin{equation}
F=-k_bT\left[ -\beta N_s(b^2-1)\bar{\lambda}+{\rm Tr}\log(-G^{-1})/\hbar\right] .
\end{equation}
In order to minimize the free energy with respect to $b$, we take the derivative,
\begin{align}
&\frac{\partial}{\partial b}\left[ {\rm Tr}\log(-G^{-1})\right] ={\rm Tr}\left[ -G\frac{\partial(-G^{-1})}{\partial b}\right] \nonumber\\
&={\rm Tr}\left[  \int\limits_{0}^{\hbar \beta}\int\limits_{0}^{\hbar \beta}d\tau d\tau'\sum\limits_{k,k'}^{}G(\tau,k,\tau',k')\dfrac{\partial}{\partial b}G^{-1}(\tau',k',\tau,k)\right]\nonumber \\
&={\rm Tr}\left[  \int\limits_{0}^{\hbar \beta}\int\limits_{0}^{\hbar \beta}d\tau d\tau'\sum\limits_{k,k'}^{}G(\tau,k,\tau',k')\left( 
\begin{array}{ll}
0 & V_k^\dagger\\V_k &- 2 b t^f_k
\end{array}
\right) \delta(k-k')\delta(\tau-\tau')
\right]\nonumber\\
&={\rm Tr}\left[ \hbar \beta \sum\limits_{k}^{}G(k)\left( 
\begin{array}{ll}
0 & V_k^\dagger\\V_k & -2 b t^f_k
\end{array}
\right)
\right],
\end{align}
where in the last step we used that $G(\tau,k,\tau',k)$ only depends on $k$ and $\tau-\tau'$. By setting $dF/db=0$, we find the first mean-field equation (\ref{eq:bequation}). Similarly, by minimizing with respect to $\lambda$, we obtain, as expected, that the second mean-field equation (\ref{eq:lequation}) is the averaged constraint condition.

Before we can use these equations to calculate the values of $b$ and $\lambda$, we need to find an expression for the Green's functions (the two-point expectation values). For this, we first investigate what happens when the Hamiltonian is diagonal in the $\Psi_k$ basis ($H_k=\sum_{k}^{}\Psi_k^\dagger D \Psi_k$, with $D$ a diagonal matrix). In that case, we find the well known Dirac distribution,
\begin{align}
\left\langle\Psi_{ik}^\dagger\Psi_{ik} \right\rangle
=\frac{{\rm{Tr}}(\Psi_{ik}^\dagger\Psi_{ik}  {\rm{e}}^{-\beta H_k})}{{\rm{Tr}}( {\rm{e}}^{-\beta H_k})}\nonumber
=\frac{\sum\limits_{n_{ik}=0}^{1} n_{ik} {\rm{e}}^{-\beta n_{ik}\epsilon_{ik}}}{\sum\limits_{n_{ik}=0}^{1}  {\rm{e}}^{-\beta n_{ik}\epsilon_{ik}}}\nonumber
=\frac{1}{1+{\rm{e}}^{\beta \epsilon_{ik}}}.
\end{align}
However, the effective Hamiltonian is generally not diagonal in (pseudo)-spin space, and we need to apply a change of basis. If $H_k$ is not diagonal in some basis $C_k$, we can relate this basis to another basis, $\Psi_k$, in which the Hamiltonian is diagonal, by the unitary matrix of eigenvectors $S_k$: $C_k=S_k\Psi_k$.
For the elements of $C_k$, this means that $C_{ki}=\sum\limits_{s}^{}S_{is k}\Psi_{s k}$.
Thus, we find that for a non diagonal basis $C_k$,
\begin{align}
\left\langle C_{jk}^\dagger C_{ik} \right\rangle=\left\langle \sum\limits_{ss'}^{}S_{js k}^*\Psi_{s_k}^\dagger S_{is' k}\Psi_{s'k} \right\rangle \nonumber
=\sum\limits_{s }^{}S_{js k}^*S_{is k}\left\langle \Psi_{s k}^\dagger \Psi_{s k} \right\rangle \nonumber
=\sum\limits_{s }^{}S_{js k}^*S_{is k}\frac{1}{1+{\rm{e}}^{\beta \epsilon_{s k}}},
\end{align}
where we used that $\left\langle \Psi_{s k}^\dagger \Psi_{s' k} \right\rangle=\left\langle \Psi_{s k}^\dagger \Psi_{s k} \right\rangle \delta_{s,s'}$. Thus, if we have a Bloch Hamiltonian $H_k$, we can now calculate the Green's function from the eigenvalues and eigenvectors of $H_k$. \cite{baruselli2014scanning2} These steps finally allow us to numerically evaluate all parts of the mean-field equations, reducing their solution to a numerical problem.

The solutions to the mean-field equations were calculated iteratively using a momentum space grid of $11^3$ points. In each step of the iteration, the values of $\lambda$, $\mu$ and $b$ were adapted accordingly, with respect to the iterative equations, 
\begin{align}
\label{eq:munum}
\mu_{n+1}&=MG(\lambda_n,\mu_n,b_n),\nonumber
\\
b_{n+1}&=\sqrt{1-\frac{1}{N_s}\sum_{k,s}^{}\left\langle f^\dagger_{ks} f_{ks}\right\rangle (\lambda_{n},\mu_{n+1},b_n)},\nonumber\\
\lambda_{n+1}&=\frac{1}{b_{n+1} N_s}\left( \sum\limits_{k}^{} {\rm Tr} \left[ \left\langle f^\dagger_{k} c_{k}\right\rangle(\lambda_n,\mu_{n+1},b_{n+1}) V_k\right] +
b_{n+1}\sum\limits_{k}^{}{\rm Tr}\left[ \left\langle f^\dagger_{k} f_{k}\right\rangle (\lambda_n,\mu_{n+1},b_{n+1})  t_k\right]\right), 
\end{align}
where $MG$ calculates $\mu_{n+1}$, such that the center of the gap is at the Fermi energy for the system defined by $\lambda$, $\mu$, and $b$, and the other equations are derived from  the mean-field Eqs.~(\ref{eq:bequation}) and (\ref{eq:lequation}).

\end{document}